\documentclass[letter,longauth,traditabstract]{aa} % for the long lists of affiliations 
\usepackage{graphicx}
\begin{document}
   \title{In-flight calibration of the $Herschel$-SPIRE instrument\thanks{Herschel
          is an ESA space observatory with science instruments provided
          by European-led Principal Investigator consortia and
          with participation from NASA.}}
   
  \author{B.\ M.\ Swinyard\inst{1} \and
        P.\ Ade\inst{2}  \and
        J-P.\ Baluteau\inst{3} \and
        H.\ Aussel \inst{16} \and
        M.\ J.\ Barlow\inst{4} \and
        G.\ J.\ Bendo\inst{5} \and
        D.\ Benielli\inst{3} \and
        J.\ Bock\inst{6} \and
        D.\ Brisbin\inst{7} \and
        A.\ Conley\inst{8} \and
        L.\ Conversi\inst{9} \and
        A.\ Dowell\inst{1} \and
        D.\ Dowell\inst{6} \and
        M.\ Ferlet\inst{1} \and
        T.\ Fulton\inst{10} \and
        J.\ Glenn\inst{11} \and
        A.\ Glauser\inst{19,20} \and
        D.\ Griffin\inst{1} \and
        M.\ Griffin\inst{2} \and
        S.\ Guest\inst{1} \and
        P.\ Imhof\inst{10} \and
        K.\ Isaak\inst{2} \and
        S.\ Jones\inst{14} \and
        K.\ King\inst{1} \and
        S.\ Leeks\inst{1} \and
        L.\ Levenson\inst{12} \and
        T.\ L.\ Lim\inst{1} \and
        N.\ Lu\inst{13} \and
        G.\ Makiwa\inst{14} \and
        D.\ Naylor\inst{14} \and
        H.\ Nguyen\inst{6} \and
        S.\ Oliver\inst{15} \and
        P.\ Panuzzo\inst{16} \and
        A.\ Papageorgiou\inst{2} \and
        C.\ Pearson\inst{1,14} \and
        M.\ Pohlen\inst{2} \and
        E.\ Polehampton\inst{1,14} \and
        D.\ Pouliquen\inst{3} \and
        D.\ Rigopoulou\inst{1,21} \and
        S.\ Ronayette\inst{1,16}\and
        H.\ Roussel\inst{17} \and
        A.\ Rykala\inst{2} \and
        G.\ Savini\inst{4} \and
        B.\ Schulz\inst{13} \and
        A.\ Schwartz\inst{13} \and
        D.\ Shupe\inst{13} \and
        B.\ Sibthorpe\inst{19} \and
        S.\ Sidher\inst{1} \and
        A.\ J.\ Smith\inst{15} \and
        L.\ Spencer\inst{2} \and
        M.\ Trichas\inst{5} \and
        H.\ Triou\inst{16} \and
        I.\ Valtchanov\inst{9} \and
        R.\ Wesson\inst{4} \and
        A.\ Woodcraft\inst{18} \and
        C. K.\ Xu\inst{13} \and
        M.\ Zemcov\inst{11} \and
        L.\ Zhang\inst{13}}

   \date{Received \today; accepted}
   
% \abstract{}{}{}{}{} 
% 5 {} token are mandatory
 
  \abstract{{SPIRE, the Spectral and Photometric Imaging Receiver, is the \it{Herschel}\rm~ {Space Observatory`s submillimetre camera and spectrometer.  It contains a three-band imaging photometer operating at 250, 350 and 500 $\mu$m, and an imaging Fourier transform spectrometer (FTS) covering 194-671 $\mu$m (447-1550 GHz). 
In this paper we describe the initial  approach taken to the absolute calibration of the SPIRE instrument using a combination of the emission from the \it{Herschel}\rm~ {telescope itself and the modelled continuum emission from solar system objects and other astronomical targets.  We present the photometric, spectroscopic and spatial accuracy that is obtainable in data processed through the ``standard'' pipelines.
The overall photometric accuracy at this stage of the mission is estimated as 15\% for the photometer and between 15 and 50\% for the spectrometer.  However, there remain issues with the photometric accuracy of the spectra of low flux sources in the longest wavelength part of the SPIRE spectrometer band.  The spectrometer wavelength accuracy is determined to be better than 1/10$^{th}$ of the line FWHM. The astrometric accuracy in SPIRE maps is found to be 2 arcsec when the latest calibration data are used.
The photometric calibration of the SPIRE instrument is currently determined by a combination of uncertainties in the model spectra of the astronomical standards and the data processing methods employed for map and spectrum calibration.  Improvements in processing techniques and a better understanding of the instrument performance will lead to the final calibration accuracy of SPIRE being determined only by uncertainties in the models of astronomical standards.}}}}


\maketitle
%
%________________________________________________________________

\section{Introduction}
The in-flight calibration of any astronomical instrument relies on a combination of previous measurements using instrumentation with known or traceable calibration accuracy, accurate models of standard astronomical sources and calibration sources internal to the instrument or facility. In this paper we discuss the methods and source models employed to convert SPIRE data to physical units and discuss the accuracy of the calibration and any caveats that must be placed on the data.

The SPIRE instrument and its overall calibration scheme are described in detail in Griffin et al. (\cite{griffin10}).  For the purposes of discussion of the calibration and performance estimation we treat SPIRE as two separate instruments that are spatially separated at the focal plane of the \it{Herschel}\rm~ telescope: a three band imaging photometric camera with bands centred nominally on 250 {$\mu$}m\ (PSW), 350 {$\mu$}m\ (PMW), and 500 {$\mu$}m\ (PLW) with detector arrays of 139, 88 and 43 respectively feedhorn coupled NTD-bolometers (Turner et al \cite{turner01}), and a two band imaging Fourier transform spectrometer (FTS) covering 194-313$\mu$m\ (SSW) and 303-671$\mu$m\ (SLW) with arrays of 37 and 19 detectors. SPIRE contains two internal calibration sources which were designed to provide a rapidly modulated signal for relative gain calibration (the PCAL source) and a stable thermal source for calibration of the spectrometer and to balance the power from the telescope (the SCAL source). PCAL is placed at an image of the telescope secondary in the common light path of the two instruments and therefore can be used to stimulate all detectors in the photometer and spectrometer. The SCAL source is placed in the second input port of the spectrometer (Swinyard et al. \cite{swinyard03}). It is therefore constantly viewed during observations and provides a reference against which the spectrum from the telescope port is measured. The intention before flight was that the SCAL source would be heated to a temperature sufficient to entirely ``null'' the spectrum from the telescope.  We discuss in Sect. \ref{spectro_pipe} why this has proved to be unnecessary for standard observing conditions.  The internal calibrators are only viewed through part of the optics chain and the overall astronomical calibration and performance of the instrument and telescope together can only be established by observation of astronomical sources as we describe in this paper. 

The structure of the paper is as follows: in Sect. \ref{initial} we describe the initial in-flight performance evaluation of the instrument before going on to cover the basic conversion of the data from digital encoder values to physical units for the detectors, the beam steering mirror (BSM) and the spectrometer mechanism (SMEC) in Sect. \ref{dataproc}.  In Sect. \ref{time_constant} we describe the characterisation of the time response of the photometer bolometers using ionising radiation hits and in Sects. \ref{photo_pipe} and \ref{spectro_pipe} we discuss how the photometric calibration was established for the photometer and spectrometer.  In Sect. \ref{testing_models} we present the results of comparing SPIRE observations to solar system objects with known or well constrained model flux densities.  The beamsize and astrometric accuracy are discussed in Sect. \ref{beam} and the wavelength accuracy in Sect. \ref{wavelength_calibration}.  We summarise the present calibration status of SPIRE and draw some conclusions for future calibration activities in Sect. \ref{conclusions}.

\section{Initial performance evaluation}
\label{initial}
The initial performance verification and comparison between in-flight and ground performance was carried out using the PCAL source whilst the \it{Herschel}\rm~cryostat lid was closed. This allowed a direct measurement of the detector response under conditions similar to those seen during ground testing. The response measured using PCAL before opening of the cryostat lid showed a close correlation between ground and flight in all detector channels in both the photometer and spectrometer. The overall in-flight measurement showed a slightly higher response in all arrays as the bolometers are approximately 10 mK colder in flight than during spacecraft level ground tests. The time response of the detector channels is a combination of the thermal response of the bolometers and the low pass filters in the readout electronics. In the photometer channels the time response is dominated by the low pass filters which have a nominal 5 Hz 3 dB cut off.  Here the PCAL flashes, which have switch-on time constant of ~90 msec and a switch-off time constant of ~50 msec, can be used to make a first order assessment of the response of the system and look for anomalous behaviour. This test is not possible for the faster (25 Hz) spectrometer filters and bolometers although analysis of the interferograms does provide a first order check.  This is achieved by measuring the difference in recorded position of the peak in signal at zero path difference when the spectrometer mechanism is scanned in its ``forward'' and ``reverse'' directions.  In both the photometer and spectrometer all channels that were working nominally before flight showed the same or similar time response. A more detailed investigation of the photometer time response was carried out using an analysis of ionising radiation glitches and is discussed in Sect. \ref{time_constant}.

Once the lid was opened we were able to observe both astronomical sources with known or well constrained model flux densities and the emission from the \it{Herschel}\rm~ telescope.  The telescope primary and secondary mirror temperatures are measured using several thermistors with a notional 0.1\% accuracy.  This, in addition to measurements of the mirror emissivity (Fischer et al. \cite{fisher04}), allows us to model the direct emission from the telescope which fills the field of view of all detectors. By varying the electrical power applied to the bolometers through changing the bias voltage - i.e. taking a so called ``loadcurve'' -  it is possible to make a direct estimate of the optical power absorbed by the bolometer. Comparing this measurement to the expected absorbed power using the modelled and ground based measurement of the instrument throughput and transmission showed that a) there is little or no straylight in the SPIRE beam due to sources on the satellite and b) that the instrument throughput and transmission are as good or better than expected before flight in both the photometer and spectrometer.

To assess the optical performance of the system and to calibrate the response of the photometer to point sources two separate types of scanned observations were carried out.  The first takes a scan map observation using the standard Astronomical Observation Template (AOT) of a point source of known or modelled flux density.  In reconstructing the map from these data the time series from many different detectors in the array are combined to make a single image of the beam.  The second approach was to scan a bright point source along the length of the array in a raster pattern with fine steps between the scan legs allowing the beam map, relative position and response of each individual bolometer to be determined.  The results from both the photometer and spectrometer arrays confirmed that the beam sizes are as expected and that no significant change to the optical alignment of the instrument had occurred during launch.  These measurements are discussed in further detail in Sect.~\ref{beam}.

\section{Data processing and calibration}
\label{dataproc}
The data processing pipeline for the SPIRE photometer and spectrometer data are discussed in detail in Griffin et al. (\cite{griffin08}) and Fulton et al. (\cite{fulton08}). In this section we highlight those elements of the processing that require the application of calibration data and discuss the origin and accuracy of the applied conversion factors.

The basic conversion from digitised output from the SPIRE electronics to voltage (in the case of the detectors) or position (in the case of the BSM and SMEC) relies on the accurate measurement of the gains and offsets of the amplification chains and encoders in the SPIRE warm electronics. These were extensively tested during the ground calibration and again in flight. Fixed resistors mounted on the bolometer arrays were used to check the detector conditioning electronics gains and no change was found compared to the ground measurement. The values of the gains applied in pipeline processing can be assumed to much more accurate than the overall calibration based on astronomical flux densities and need not be considered in estimating the calibration accuracy.

The SMEC position encoder accuracy was tested by comparing the measured and expected frequencies of known lines (see Sect. \ref{wavelength_calibration}).  The position encoder of the BSM has been recalibrated in flight by using it to move a bright point source across the field of view and using the known positions of the detectors (see Sect. \ref{beam}) to determine its position.  Using this method, combined with an end to end optical model, we can determine the position of the BSM to within an equivalent of 0.5 arcsec on the sky, much better than the absolute pointing error of the \it{Herschel}\rm~ satellite which is estimated as 2 arcsec (Pilbratt et al. \cite{pilbratt10}).

\section{Photometer bolometer time constants}
\label{time_constant}
The basic time response of the photometer detector channels is dominated by the low pass filtering in the signal conditioning electronics which has been accurately modelled and verified during ground testing. However, the thermal response of the bolometers themselves has a second order effect that varies with bolometer temperature and it is important that this is understood as correcting for it is critical in both the astrometric and photometric accuracy achievable in SPIRE maps. The first assumption has been that the bolometer thermal response is represented by an 'RC' type response characterised by a single time constant. Investigation into the presence of any slower bolometer time constants on scan map and spectrometer time domain data has so far revealed no evidence for a slow response to optical power and only the correction for a first order fast time response is currently being performed by the pipeline processing. 

High energy particles hitting the bolometers cause signal glitches as the energy is absorbed and the temperature of the bolometer increases. The response to these glitches gives a direct measurement of the bolometer and electronics time constants. To measure the photometer bolometer time-constant the data from a large number of photometer glitches have been collected and compared to the theoretical impulse response modelled as a delta-function convolved with the photometer electronics and bolometer transfer functions. Applying this process to sets of individual (single event) glitches, reveals that the photometer glitch response for most photometer bolometers can be best described by a two-component transfer function, with a fast time constant $\tau_1$ $\approx$ 6 ms and a second slow component with a time constant of $\tau_2$ $\approx$105 ms and amplitude about 6.6 \% (Fig. 1).  A second type of common, concurrent glitches are also seen which are most likely due to interaction between particles and the bolometer array substrate.  Applying the same process to these reveals that the common glitch response is best described by a first-order-only bolometer transfer function with a time constant of about 6 ms.  This disparity in the system response between single-event and common glitches, and how this relates to the absence of any slow response to optical power, is not yet understood and is currently being investigated. The frequency response correction of both the electronics and the bolometers amounts to no more than 5\% of the signal for standard scan map data and given our knowledge of the electronics and bolometers any uncertainties in this correction can, again, be safely ignored in the overall calibration uncertainty.

\begin{figure}
\includegraphics[angle=0,width=80mm]{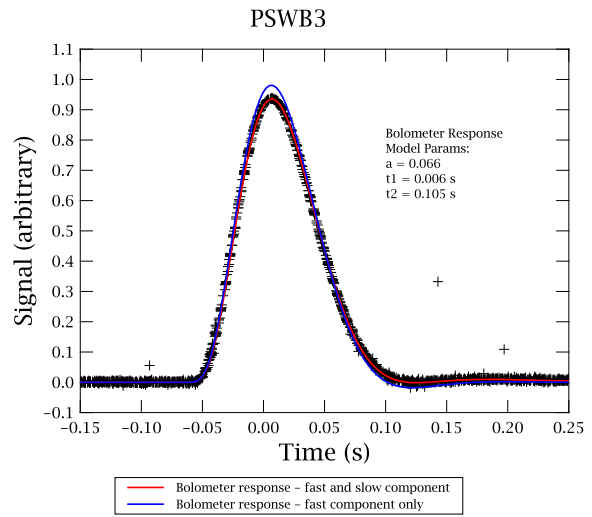}
\label{figure1}
\caption{The average response of a photometer bolometer to an ionising radiation particle derived from the co-addition of many glitches (crosses) compared to the modelled impulse response of the electronics and bolometer with (red) and without (blue) a slow time response component (see text).}
\end{figure}

\section{Photometer photometric calibration}
\label{photo_pipe}
The conversion from voltage to flux density in the photometer is based on the assumption that the small signal responsivity can be related to the bolometer operating point voltage nearly linearly i.e. that we can write the inverse responsivity in the form
\begin{equation}
\frac{dS}{dV} = K_1 + \frac{K_2}{V-K_3}
\label{equation1}
\end{equation}
where S is the flux density, $K_1,K_2,K_3$  are constants and $V$ is the measured voltage across the bolometer.  $K_1$ is in units of Jy/V, $K_2$ is in Jy and $K_3$ is in V.  Integrating this equation gives
\begin{equation} 
S = K_1(V-V_0) + K_2 \ln \frac{[V-K_3]}{[V_0-K_3]}
\label{equation_calibration}
\end{equation}
where now $V_0$ is another arbitrary constant that we chose to represent the bolometer voltage measured under a standard observing condition on a dark region of the sky.  The various parameters have been determined from measurements in-flight using PCAL to measure the relative responsivity while viewing sources of different brightnesses in order to change the bolometer voltage.

\begin{figure} 
\centering
   \includegraphics[angle=0,width=80mm]{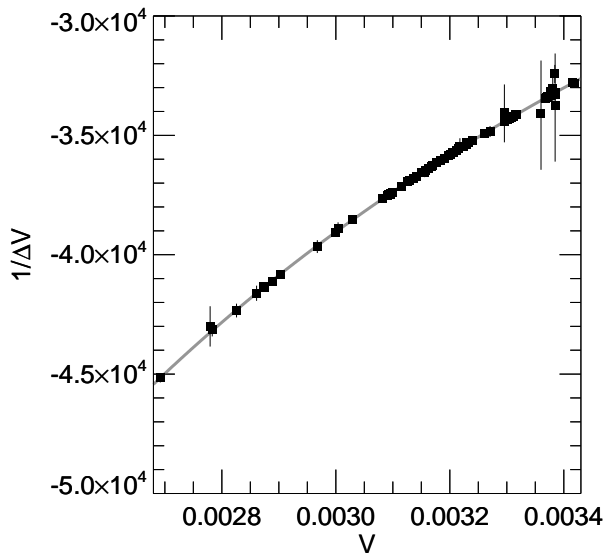}
\caption{An illustration of the basic method of photometric calibration.  Here we plot the response to the internal calibrator ($1/\Delta V_{PCAL}$) versus the bolometer operating point voltage ($V$) as measured using flux densities ranging from dark sky to the peak of SgrA by pointing at different locations in the region of SgrA.}
\label{figure_calibration}
\end{figure}

The calibration curves were measured for each detector in two steps.  First, the change in voltage $\Delta V_{PCAL}$ observed for PCAL flashes were measured when the telescope was pointed at regions of the sky with different flux densities ranging from dark sky to the peak of Sgr A (for the nominal voltage settings) or Sgr B2 (for the lower gain bright source mode settings).  The plot of $1/\Delta V_{PCAL}$ versus $V$ gives the unscaled version of Eq.~\ref{equation1}; an example is shown in Fig.~\ref{figure_calibration} where the telescope was pointed at the dark sky and different locations in the region of SgrA to vary the flux onto the detectors.  The calibration curve itself is then scaled using data from observations in which Neptune (or a source calibrated using Neptune) is scanned across each detector in each array.  The feedhorns and bolometers are not inherently sensitive to polarisation and any instrinsic polarisation in the calibration targets will have negligible effect on the overall photometric calibration.

The initial absolute calibration scale was set using the asteroid Ceres and a model of its submillimetre flux density based on the Standard Thermal Model (Lebofsky et al \cite{lebofsky86}).  We have since observed Neptune and correction factors between the output of the photometer pipeline and the model of Neptune based on the work of Moreno (\cite{moreno98}, \cite{moreno10}) have been established (see Griffin et al. \cite{griffin10}).   We discuss the accuracy of the Moreno model further in Sect. \ref{testing_models} and the accuracy of the calibration constants in Sect. \ref{conclusions} of the present paper.

\section{Spectrometer pipeline photometric calibration}
\label{spectro_pipe}
The calibration of the spectrometer follows a different method compared to the photometer.  The signal that is measured by the spectrometer detectors is not a direct measurement of the flux density integrated over the passband as in the photometer but rather the Fourier component of the spectral content.  Therefore Eq. ~\ref{equation1} is not directly applicable but an analogous equation can be used to correct for any non-linearity between absorbed power and bolometer voltage before transforming into the frequency domain.  The parameters in this scheme are derived from a model response of the bolometer and have different values depending on the bolometer bias that is set.

Once a linearised timeline in volts has been obtained the signal versus optical path difference is calculated using the mechanism position and further corrections for phase error are made (see Fulton et al. \cite{fulton08}).  The source, telescope, SCAL and the instrument self emission are always measured together and it is necessary to subtract a reference interferogram taken on dark sky to obtain the source on its own.  This difference interferogram is transformed into spectral space and finally converted to flux density using a relative spectral response function (RSRF).  The RSRF is derived by taking the interferogram of an astronomical source with a well modelled continuum, subtracting the reference interferogram, transforming this into spectral space and dividing by a model of the source spectrum.  In the present pipeline we use the asteroid Vesta as the calibration source with a continuum model provided by T. Mueller (Mueller and Lagerros, \cite{mueller02}).

Interferograms taken on Vesta, Neptune and Uranus showed that, with the SCAL source off, the signal at the central peak does not saturate, or at most only a few samples are saturated, once the detector bias is correctly set.  The fact that we do not require SCAL to null the telescope emission is a consequence of the lower total emission from the telescope and straylight compared to the expected values used in the initial design of the SPIRE instrument.  Given this, and that using SCAL adds to the photon noise in the measurement, we have decided not to use the SCAL source in routine observations.  An additional benefit of this mode of operation is that, with SCAL off it, and the rest of the instrument, are at a temperature between 4.5 and 5 K and the thermal emission from these components is limited to frequencies only detectable in the SLW band.

The standard calibration for SPIRE spectra is based on a point source.  An alternative calibration can be derived using the telescope which is appropriate for a source that entirely fills the detector field of view.  A correction based on measurements of the spectrometer beam shape is required to convert from one to the other (see Sect. \ref{testing_models}).  It should also be noted that instrument dependent spectral features in the passband of the spectrometer will change as a function of source extent.
\section{Testing the Astronomical Models}
\label{testing_models}
\begin{figure}
\includegraphics[angle=0,width=80mm]{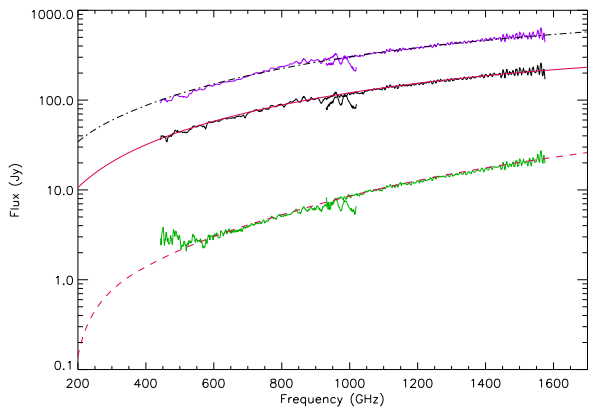}
\caption{The spectra of three solar system objects observed using the SPIRE Fourier transform spectrometer.  From bottom to top we plot Vesta, Neptune and Uranus.  The conversion to flux density was achieved using the \it{Herschel}\rm~ telescope as a calibration standard as described in the text.  We plot the model spectra of the three sources from respectively Moreno (\cite{moreno98}, \cite{moreno10}) for Uranus and Neptune and Mueller (\cite{mueller02}) for Vesta.  The Uranus model spectrum has been increased by 10 Jy at all wavelengths and the Vesta model has been multiplied by a factor of 1.2.}
\label{spectra}
\end{figure}

To test the accuracy of the models used for both the spectrometer and photometer we have derived the calibration using an alternative method where we use the telescope itself plus knowledge of the instrument throughput (see Sect. \ref{beam}).  As discussed in Sect. \ref{initial} the telescope is represented by a blackbody with a well known relative dependence between emissivity and wavelength even if the absolute overall emissivity may have an uncertainty up to 30\% (see Fischer et al. \cite{fisher04}).  We can therefore generate an RSRF from the measured telescope spectrum on dark sky and use this to calibrate the spectra obtained on Vesta, Neptune and Uranus and compare to the models to check for self consistency.  The results together with continuum models are shown in Fig. \ref{spectra}. The basic comparison is with the Neptune model of Moreno which shows an excellent agreement across all wavelengths giving confidence in both the relative and absolute level of the telescope emissivity.  The Uranus model shown is also that of Moreno, but here we have increased the continuum by 10 Jy at all frequencies to give a better fit to the spectrum calibrated from the telescope. Reapplying a calibration file derived from this model of Uranus to Neptune shows that the Neptune measurement agrees with the model to within 5-10\% across the full SPIRE band.  The Vesta model as provided by Mueller was shown to be ~15-20\% lower than the measurement calibrated using Neptune and so we have increased the model spectrum  that is used in the derivation of the spectrometer pipeline calibration by 1.2.  We discuss the accuracy of the pipeline derived spectra further in Sect. \ref{conclusions}.

 \section{Beam and astrometric calibration}
\label{beam}
\subsection{Photometer}
\begin{figure}
\includegraphics[angle=0,width=80mm]{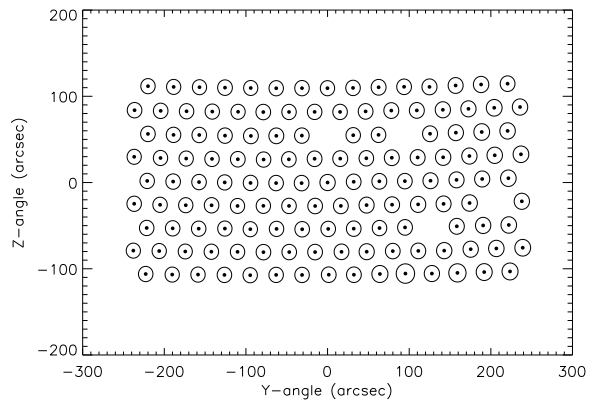}
\caption{A representation of the position (centres of the circles) and beam size (radius of the circles) for the photometer PSW detector array.  The missing circles indicate non-operational detectors.}
\label{phot_beams}
\end{figure}

The measurement of the beam size of the photometer has been discussed in Sect. \ref{initial}.  The average beam derived for the three arrays has a near Gaussian core with a width of 18.1, 25.2 and 36.6 arcsec respectively all with an uncertainty of $\pm$5\%.  However, the full PSF outside of this core has much structure due to the diffraction from the secondary mirror and its support structure and it is recommended to use equivalent beam areas of 501, 944 and 1924 arcsec$^2$ respectively to calculate surface brightness when observing extended sources (SPIRE Observer's Manual \cite{spire10}). These values are quoted as the mean for all detectors in an array, there is some minor variation across the arrays as demonstrated in Fig. \ref{phot_beams} which shows the centroid positions and the relative beam sizes from a point source scanned across the PSW array. The variation in beam size has been  found to vary  by no more than 10\% for the majority of detectors across any of the three arrays.  In the example shown here the maximum and minimum beam sizes of the nominally operational detectors are ~17.3 and ~19.5 arcsec.  The apparently large beamsizes in two of the detectors are caused by their very slow time response and these detectors are excluded from the map making.  Measurement of the beam width versus wavelength
within the bands is not possible in flight but characterisation was performed during ground test campaigns (Ferlet et al. \cite{ferlet08}).  

The measured positions of the detectors in each array are folded into the map reconstruction algorithms together with the knowledge of the spacecraft pointing. Comparison between fields with sources seen at other wavelengths (mostly from \it{Spitzer}\rm~and radio catalogues) shows that, with the present knowledge and algorithms, the typical astrometric accuracy of SPIRE maps processed with the current version of the \it{Herschel}\rm~processing pipeline is around 4 arcsec. The cause of part of the inaccuracy has already been identified and subsequent versions of the pipeline have an astrometric accuracy of 2 arcsec or better.  This should be compared to the telescope relative pointing error of 0.2 arcsec 1-sigma over 1 minute and the spatial relative pointing error of 1.5 arcec (Pilbratt et al. \cite{pilbratt10}).  

\subsection{Spectrometer}
\begin{figure}
\includegraphics[angle=0,width=80mm]{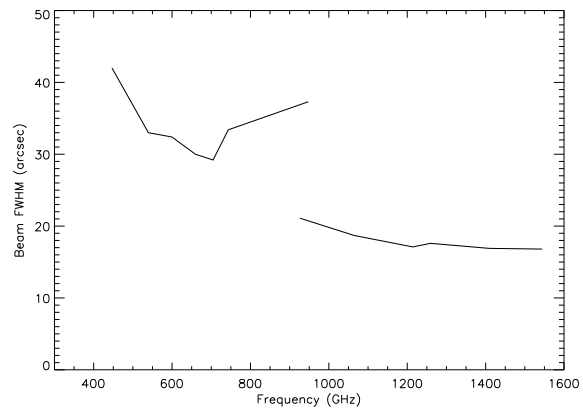}
\caption{The measured spectrometer beam width as a function of wavenumber for the central detectors in the two arrays.}
\label{spectrometer_beam}
\end{figure}

The shape and extent of the beam in the spectrometer has been subject to much study (Ferlet et al. \cite{ferlet08}) both on the ground and in flight. The beam size versus frequency has been measured directly by taking medium resolution spectra on a point source of Neptune placed at different locations with respect to the central detectors by stepping the satellite position. The results are shown in Fig. \ref{spectrometer_beam}. The highly structured nature of the variation in beam size with frequency is expected from the multi-moded feedhorns used for the spectrometer arrays. The beams are only Gaussian at the low frequency end of each band and methods for conversion between calibration valid for point and extended sources require further investigation.

\section{Wavelength calibration}
\label{wavelength_calibration}
\subsection{Photometer}
The central wavelengths for the three photometer bands are derived from pre-flight measurements of the passbands of the detectors and the instrument optical filter chain (Spencer \cite{spencer09}, SPIRE Observers Manual \cite{spire10}).  As described in Griffin et al. (\cite{griffin10}) these are integrated over the waveband assuming a flat spectrum to give band centres at 250, 352 and 504 {$\mu$}m.  The edges of the bands as defined at 50\% of the average in band transmission are 211-290 {$\mu$}m (PSW),  297-405 {$\mu$}m (PMW), and 409-611 {$\mu$}m (PLW).  The filter profiles themselves are available as standard calibration products within the \it{Herschel}\rm~ processing environment. 

\subsection{Spectrometer}
The accuracy of the wavelength calibration was derived from line fits to $^{12}$CO lines in the spectra of 5 sources: Orion Bar, CRL618, NGC7023, S106 and DR21. The spacecraft velocity and the source velocity were subtracted from the fitted line positions to give the residual velocity offset.   
%Figure \ref{CO} shows the residual differences between the expected and measured line centres.  The filled circles in the plot give the average offset over all sources, and the error bars show the standard deviation. 
Using this method we found a calibration accuracy within 1/10th of a resolution element across the spectrometer band as defined by the FWHM of the Sinc profile - i.e. 1.207$\times 0.04$~cm$^{-1}$ - and that the average line centre offset was 27 km s$^{-1}$. Much of this offset can be explained by the detailed physics of the interferometer (Spencer et al. \cite{spencer10}) which is now understood and we expect that all instrumental offsets will be removed for all detectors across the field of view in future pipeline processing.  The two spectrometer channels' transmission profiles fall rapidly at the edges and we define the trustworthy  band edges for the central detectors of the two channels to be 14.9-33.0 cm$^{-1}$ and 32.0-51.5 cm$^{-1}$.  These edges vary somewhat depending on the position of detectors across the field of view but nowhere do they change by more than 0.2 cm$^{-1}$.

%\begin{figure}
%\includegraphics[angle=0,width=80mm]{special_issue_wave_accuracy.ps}
%\caption{The average observed velocity error versus line frequency derived from $^{12}$CO transitions in five sources using the %SPIRE Fourier Transform Spectrometer. The error bars indicate the spread in the measured line centres and the dashed line %indicates 1/10th of the line FWHM (see text).}
%\label{CO}
%\end{figure}

\section{Discussion and conclusions}
\label{conclusions}
The most significant part of calibrating an astronomical instrument is the accurate conversion from measured signal to source flux density. In this paper we have given an overview of how this is currently carried out for the \it{Herschel}\rm-SPIRE instrument.  We have demonstrated an inter-comparison between the models used for astronomical sources and the \it{Herschel}\rm~telescope emission spectrum that gives a final agreement between the Neptune model used as the primary calibration standard and our measured spectrum of between 5-10\%.

The calibrator source model is only one component of the final uncertainty and we have tested the conversion parameters for the photometer Volts to Jy using a Monte Carlo approach to determine the variance in the calibration curve as a function of voltage. This method involved creating multiple calibration curves of $1/\Delta V_{PCAL}$ versus $V$ with addition random noise representative of the measurement errors and measuring the effect on the uncertainty in derived calibration coefficients. This translates into uncertainties that are $\leq$ 0.1\% for all detectors.  This is much smaller than the other two sources of error, which are the uncertainty in the model for Neptune (5-10\%) and the uncertainty in peak flux density measurements for Neptune derived from the maps after processing. Taking these factors into account for data processed with the current processing software the overall photometric error can be assumed to be 15\%.

For the spectrometer photometric calibration a comparison of the pipeline derived Neptune spectrum based on Vesta to the Neptune model shows a typical variation of 15-20\% in the SSW band and 20-30\% in the SLW band above 20 cm$^{-1}$ and up to 50\% below 20 cm$^{-1}$.  It should be noted that variations in the temperature of the instrument itself between the measurement of the dark spectrum and the source can cause extra flux to be observed in the SLW band. The net result is an additive term in the spectrum below 25 cm$^{-1}$ of the order of a few Jy which can leave significant positive or negative extraneous flux density levels in the source spectrum especially for faint objects. With careful interactive processing it is possible to remove this signature and obtain accurate source flux densities (Ivison et al. \cite{ivison10}, Bocklee-Morvan et al. \cite{bockl10}). Further investigation on this aspect of the spectrometer calibration is ongoing and will be reported in future work.

The calibration of the data from the SPIRE instrument processed through the current data reduction pipeline software is already of a standard more than sufficient to allow significant scientific conclusions to be drawn from the observations.  Future work will concentrate on improving the photometry using other astronomical sources and source models, including cross calibration with other facilities such as the \it{Planck}\rm~satellite.   We will also work on improvements to the data processing tools, the astrometry of the maps through better attitude reconstruction and the wavelength calibration of the spectrometer through more sophisticated instrument performance models.

\begin{acknowledgements}
SPIRE has been developed by a consortium of institutes led by Cardiff Univ. (UK) and including Univ. Lethbridge (Canada); NAOC (China); CEA, LAM (France); IFSI, Univ. Padua (Italy); IAC (Spain); Stockholm Observatory (Sweden); Imperial College London, RAL, UCL-MSSL, UKATC, Univ. Sussex (UK); Caltech, JPL, NHSC, Univ. Colorado (USA). This development has been supported by national funding agencies: CSA (Canada); NAOC (China); CEA, CNES, CNRS (France); ASI (Italy); MCINN (Spain); SNSB (Sweden); STFC (UK); and NASA (USA)     
\end{acknowledgements}

\institute{Space Science and Technology Department,                            %1
              Rutherford Appleton Laboratory, Chilton,
              Didcot, Oxon UK\and
              Cardiff University, The Parade, Cardiff \and              %2
              LAM, Marseille, France \and                                         %3
              University College London, UK \and                                  %4
              Imperial College, London, UK \and                                   %5
              NASA JPL, Pasadena, California, USA \and                            %6
              Cornell University, Ithaca, New York, USA \and                      %7
              Department of Astronomy and Astrophysics, University of Toronto,    %8
              Toronto, Canada \and
              European Space Agency (ESA),                                        %9
              European Space Astronomy Centre (ESAC),
              Villanueva de la Ca\~nada,
              Madrid, Spain \and
              Blue Sky Spectroscopy Inc., Lethbridge, Canada \and                 % 10
              University of Colorado, Boulder, Colorado, USA \and                 % 11
              Caltech, Pasadena, California, USA \and                             % 12
              NASA Herschel Science Centre, IPACS, Pasadena, California, USA \and % 13
              Institute for Space Imaging Science, University of Lethbridge, Canada\and  %14
              Astronomy Centre, Dept. of Physics \& Astronomy, University of Sussex, UK \and                                       % 15
              CEA Saclay, Gif-sur-Yvette, France \and                                             % 16
              Institut d'Astrophysique de Paris, Universit\'e Pierre \& Marie Curie,
   98 bis Boulevard Arago, Paris, France \and                                             % 17
              SUPA, Institute for Astronomy, University of Edinburgh, Blackford Hill, Edinburgh, UK\and %18
              UK Astronomy Technology Centre, Blackford Hill, Edinburgh, UK \and % 19
              Institute of Astronomy, ETH Zurich, 8093 Zurich, Switzerland \and %20
              Department of Physics, Oxford University, Keble Road, Oxford, UK}  %21

\end{document}